\ifx \doiurl    \undefined \def \doiurl#1{\href{http://dx.doi.org/#1}{\textsf{DOI}}}\fi
\ifx \adsurl    \undefined \def \adsurl#1{\href{http://adsabs.harvard.edu/abs/#1}{\textsf{ADS}}}\fi
\ifx \arxivurl  \undefined \def \arxivurl#1{\href{http://arxiv.org/abs/#1}{\textsf{arXiv}}}\fi
\documentclass[pdftex,12pt]{article}
\usepackage{natbib}
\usepackage[T1]{fontenc}
\usepackage[english]{babel}
\usepackage[breaklinks,colorlinks]{hyperref}
\hypersetup{
citecolor=blue,
urlcolor=blue
}
\usepackage{setspace}
\usepackage{graphicx}        % For eps figures, newer & more powerfull
\usepackage{color}           % For color text: \color command
\usepackage{breakurl}        % For breaking URLs easily trough lines
\def\arcsec{$^{\prime\prime}$}
\usepackage[labelsep=period]{caption}

\usepackage{textcomp}
\usepackage{gensymb}
\usepackage[font={small}]{caption}
\title{\vskip-2cm
Peculiarity of the oscillation stratification in sunspot penumbrae}

\author{D.Y.~Kolobov, A.A.~Chelpanov and N.I.~Kobanov\\
    \small{Institute of Solar-Terrestrial Physics} \\
    \small {of Siberian Branch of Russian Academy of Sciences, Irkutsk, Russia} \\
    \small {email: \url{kolobov@iszf.irk.ru}}
    }

\date{\small{[%\href{}
{This article was firstly published in {\textit{Solar Physics}} \href{https://dx.doi.org/10.1007/s11207-016-0953-7}{DOI}}]}}

\begin{document}
\maketitle
\begin{abstract}
Spatial distributions of the dominant oscillation frequency obtained for four
sunspots show a feature shared by all the analysed levels of the solar
atmosphere in these sunspots. This feature located in the inner penumbrae indicates
that this region has favourable conditions for 2.5--4\,mHz oscillation
propagation. This agrees with the fact that the spectral composition of the
oscillations at three atmospheric heights (Fe\,\textsc{i}\,6173\,\AA, 1700\,\AA,
and He\,\textsc{ii}\,304\,\AA) in this region are similar.
There have been previous evidence of particular similarities along height of
photospheric magnetic field strength, line-of-sight velocity, and temperature
profile in the inner penumbra, where the internal boundary of the Evershed flow
is located. The finding of the same dominant oscillation frequency at a range of
altitudes from the chromosphere up to the transition region extends the height
range, suggesting similarities in physical conditions.

\end{abstract}
% \keywords{Sunspots, Penumbra; Oscillations; Magnetic topology}
\section{Introduction}
     \label{S-Introduction}
\par Sunspots have long been an important object for the study of
oscillations and waves in the solar atmosphere.
Sunspots provide a diverse range of interactions between the solar magnetic field and matter.
The umbra, where the vertical magnetic field prevails, shows
signatures of downward motion and weak five-minute oscillations of the whole umbra
at the photospheric level \citep{Lites, Kobanov90}. In the chromosphere, strong
three-minute oscillations dominate in the umbra; these oscillations at first were
considered as standing acoustic waves \citep{Lites, Georgakilas}. Later, these
waves were shown to be moving upward \citep{Rouppe}; also they were shown not to
be a simple continuation of the running penumbral waves (RPW) in the horizontal
direction \citep{Kobanov06}. New observations with ever-increasing spatial and temporal resolution
capabilities reveal many new facts about the sunspot's fine structure
\citep{Jess15, Khomenko, Yuan, Sych14, Sych15}.

\par Penumbra is a complex part of a sunspot, whose understanding and modelling
are largely complicated by horizontal---or rather a mixture of differently
inclined---flows, steep temperature gradient, and rapid change of the magnetic
field strength towards the outer boundary. Further complexity arises due
to the inhomogeneity in the azimuthal direction: the bright and dark
filaments are associated with different physical parameters. The magnetic field
inclination has been shown to be larger in the dark filaments forming the
so-called uncombed penumbra \citep{Title, Solanki, BellotRubio03}. Recently,
with the help of high-resolution instruments, two components of the magnetic
field were observed in a sunspot penumbra. Both components have an inclination
close to 40\degree\ in the inner penumbra, and in the outer penumbra the
dark-filament field grow horizontal, while the bright-filament field only
reaches 60\degree\ inclination \citep{Langhans}.

\par It is necessary to analyse the oscillations in the penumbra to form a
comprehensive picture of the waves propagating in sunspots. A wide range of
frequencies is observed in sunspot penumbrae: \citet{Lites, Brisken, Zirin}
registered oscillations in intensity and Doppler velocity signals.
Different frequencies tend to
appear at different regions of the penumbra: typically, the longer the period,
the farther it is observed from the sunspot centre. \citet{Sigwarth} noted that
this change in the frequency is more pronounced in the chromosphere than in the
photosphere. Such a distribution is explained by the increase in the magnetic
field inclination closer to the boundaries of a sunspot, and thus the decrease in the
cut-off frequency \citep{Reznikova, Reznikovaetal, Kobanov13}.

\par In a sunspot there is a phenomenon called
running penumbral waves (RPW) --- observed
increase in brightness travelling outwards the penumbra \citep{Beckers, Giovanelli, Zirin}.
These waves span a range of frequencies from
1 to 4~mHz \citep{Lites}. \citet{Lites, Brisken, Jess} showed that the periods
of the waves increase closer to the boundaries of a sunspot. RPWs were
found at the photospheric heights as well \citep{Musman,Lohner}.  Naturally, the
question was raised on the origin of RPWs: two concepts were proposed.
The first one is that RPWs are real waves propagating horizontally across the penumbra
\citep{Alissandrakis, Tsiropoula, Tziotziou04, Tziotziou06}; and the second concept
implies them to be an apparent effect---a result of waves rising to the surface along
inclined magnetic tubes, thus appearing first at the inner penumbra and then
farther from the sunspot centre \citep{Rouppe, Bogdan, Kobanov06, Bloomfield,
Kobanov08,cho2015apj}. The same effect is also deemed to be responsible for umbral flashes;
and indeed, recently, researchers tend to consider umbral flashes and RPWs as
manifestations of one phenomenon \citep{Madsen}. This second explanation raises
a consequent question: what is the origin of the waves responsible for the
observed effect? A number of authors come to the conclusion that these waves are
slow magnetoacoustic modes resulting from photospheric p-mode oscillations
\citep{Bloomfield, Madsen} or a broadband energy deposition process, \textit{e.g.} granulation
motions \citep{bot2011apj}. \citet{Jess} studied the influence of the magnetic
field inclination on the RPW. They concluded that the increase in the
inclination leads to the increase in the dominant periodicity due to the
dropping of the cut-off frequency. However, due to the complicated dynamical
properties, reliable estimations of the RPW parameters are difficult to carry
out.

\par The important feature of the oscillations discussed above is their
relation to the Evershed flow.
\citet{Kobanov04a} found three ranges of oscillations that most likely connect
the direct Evershed flow and the inverse---the so-called St.~John's---flow.
The 20--35-minute oscillations have the most consistent
phase difference between the photospheric and chromospheric heights.

\par In our previous works \citep{Kobanov15,Kolobov}, we revealed that the
dominant frequency spatial distribution shows a peculiarity in sunspots'
penumbra. We constructed plots showing the dominant frequencies averaged in
the azimuthal direction as a function of the distance to sunspot's barycentre,
and these plots converged in the inner penumbra. This feature corresponds to the
3--4 mHz frequency range, which indicates that five-minute oscillations dominate
above the inner penumbra at all the heights from the photosphere to the
transition region. One can assume that in this ring-shaped region above the
inner penumbra, favourable physical conditions for five-minute wave propagation exist.

\section{Methods}

\par For this study we used full-disk narrow-band images provided by
the \textit {Atmospheric Imaging Assembly} (AIA: \citealp{lem2012sp}) onboard
the \textit{Solar Dynamics Observatory} (SDO).
Of all the spectral bands used in the observations we
chose four:  1700\,\AA, He\,\textsc{ii}\,304\,\AA, Fe\,\textsc{ix}\,171\,\AA\
obtained by AIA, and Fe\,\textsc{i}\,6173\,\AA\
obtained by \textit{Helioseismic and Magnetic Imager} (HMI: \citealp{sch2012sp}).
These lines cover the height range from the deep photosphere to the corona.

\par The AIA data have time resolution 12 and 24\,s, depending on the spectral
band. Each pixel of these data corresponds to 0.6\arcsec. The HMI data have a
45\,s time resolution and 0.5\arcsec\ detector pixel size. Fe\,\textsc{i}\,6173\,\AA\ line is formed
in the photosphere at a height of 200\,km \citep{Beckers}. Also, the Doppler
velocity maps and magnetograms are available in this line.

\par In the analysis we used four sunspots in NOAA 11311, 11479, 11711, and 12149
(Table~\ref{tbl:spots}).
In all four cases the spots had roughly regular round shapes and located close to
the disk centre, and thus the effects of the foreshortening and asymmetry were minimal.
The data series lengths ranged from 4 to 8 hours, and no flares were registered
during these observations.

\begin{table}
\caption[]{\raggedright Data series used in the analysis}
\begin{tabular}{ccrccccc}
\hline  \\
{\bf No.}&				% 1
{\bf NOAA}&				% 2
{\bf Date}&				% 3
{\bf Disc}&				% 4
{\bf T$_{start}$ (UTC)}&			% 5
{\bf T$_{end}$ (UTC)}			% 6
\\
{\bf}&					% 1
{\bf}&					% 2
{\bf}&					% 3
{\bf location}&				% 4
{\bf hh:mm}&				% 5
{\bf hh:mm}				% 6
\\ \hline
\\ 1 & 11311 &  6 Oct 2011 &  11$^\circ$S  08$^\circ$E  &  03:21 & 04:45
\\ 2 & 11479 & 16 May 2012 &  15$^\circ$N  11$^\circ$E  &  00:07 & 07:59
\\ 3 & 11711 &  6 Apr 2013 &  17$^\circ$S  03$^\circ$W  &  00:00 & 04:20
\\ 4 & 12149 & 27 Aug 2014 &  10$^\circ$N  02$^\circ$W  &  06:00 & 10:00

\\ \hline
\end{tabular}
\label{tbl:spots}
\end{table}

\par After removing trends from the time series, the power spectra were
calculated by Fast Fourier Transform (FFT) using the standard routine of the
Interactive Data Language (IDL).
Based on these spectra, we constructed
the dominant frequency distributions within the analysed regions. To this end,
the spectrum from each pixel was smoothed using convolution with a 1\,mHz
window, and the maximum frequency value of the resulting function was considered
as the dominant frequency within the given pixel.

\section{Results and Discussion} %%%%%%%%%%%%%%%%%%%%%%%%%%%%%%%%%%%%%%%%

\par Oscillations of different frequencies are observed in sunspots, and within
a sunspot they are distributed non-uniformly.
First, different frequency bands occupy different regions of a sunspot.
High-frequency oscillations tend to be located within the umbra boundaries, while
lower frequencies form circles, whose radii increase with the decreasing frequency.
Such distributions are believed to be related to the magnetic field
configuration, namely, magnetic field inclination angle: the location of high
frequencies coincides with that of the vertical magnetic field, while low frequencies
are located at the high-inclination outer penumbra \citep{Reznikova, Kobanov13}.
Second, the circles of oscillation distributions grow with the height. Such
a pattern, again, is deemed to be related to the magnetic field configuration:
waves propagating along magnetic field lines inclined from the spot centre
approach the spot's outer boundaries at each consecutive height level
\citep{Kobanov13}.

\begin{figure}
\centerline{
\includegraphics[width=10cm]{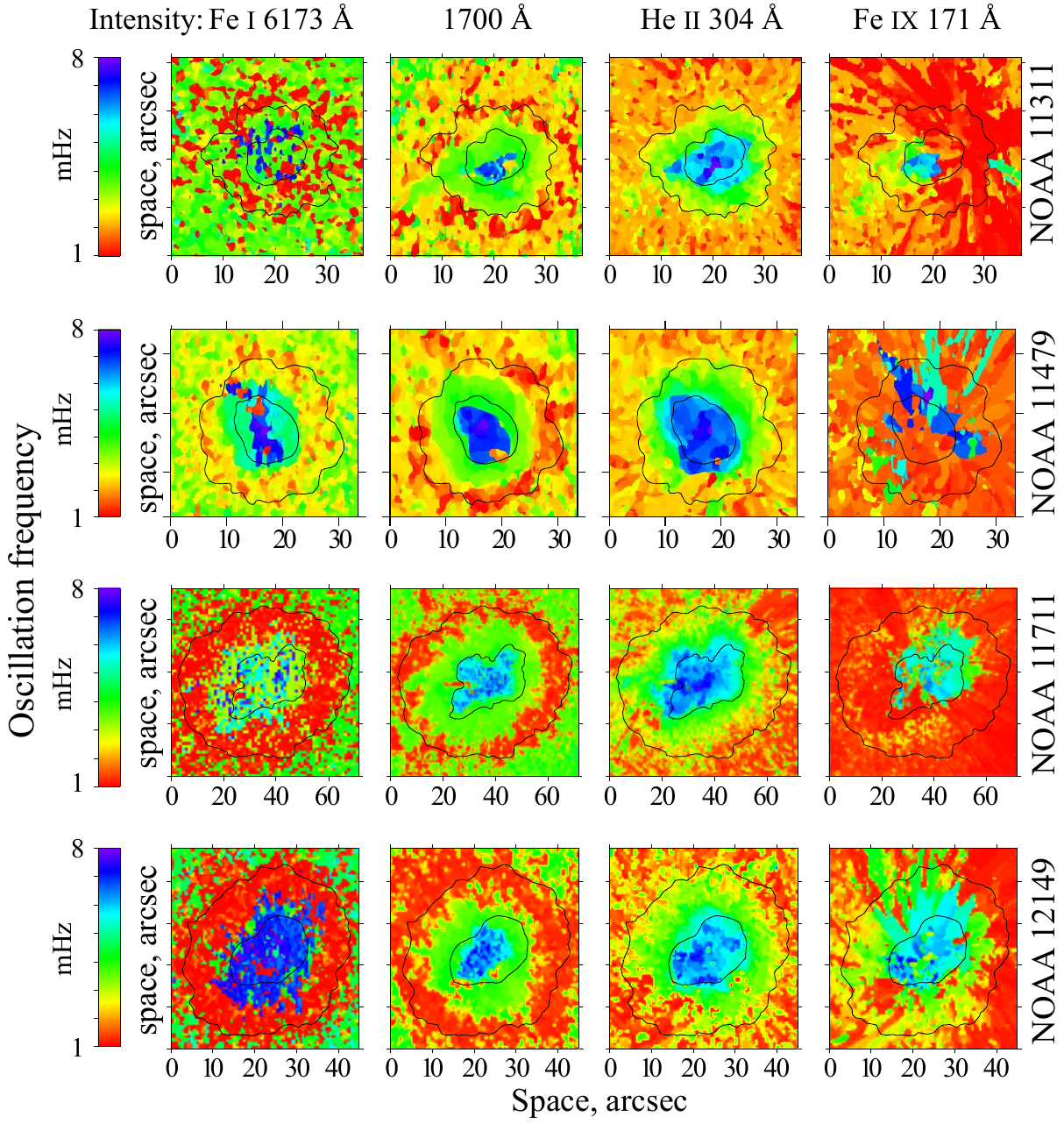}
}
\caption{Dominant frequency distributions in sunspots at different height levels,
from the photosphere to the corona. The black lines show the inner and outer
penumbra boundaries as seen in the 1700\,\AA\ band AIA images}
\label{fig:dominant}
\end{figure}

\par
\begin{figure}
\centerline{
\includegraphics[width=12.5cm]{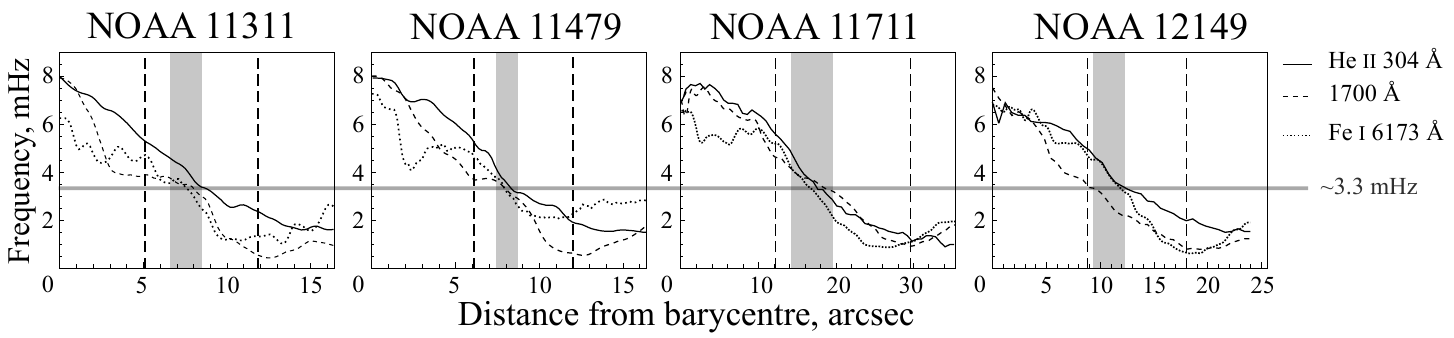}
}
\caption{Radial distributions of the dominant frequencies at three heights in
sunspots as functions of the distance from the barycentre. The vertical dashed lines
mark the penumbra boundaries. The region of interest is marked with grey area, which
corresponds to the averaging over the circular-shaped domain in the penumbra}
\label{fig:radial}
\end{figure}

 Figure\,\ref{fig:dominant} shows spatial distributions of the dominant
frequencies based on FFT power spectra. The distributions are reconstructed from the SDO data.
They follow the two aforementioned features of the oscillation behaviour in
sunspots. Based on these distributions we plotted profiles of dominant
frequencies as functions of distance from a sunspot centre \textit{r} presented
in Figure\,\ref{fig:radial}. These profiles were plotted for three height
levels---from the photosphere to the transition region---in the four sunspots.

\par
\begin{figure}
\centerline{
\includegraphics[width=13cm]{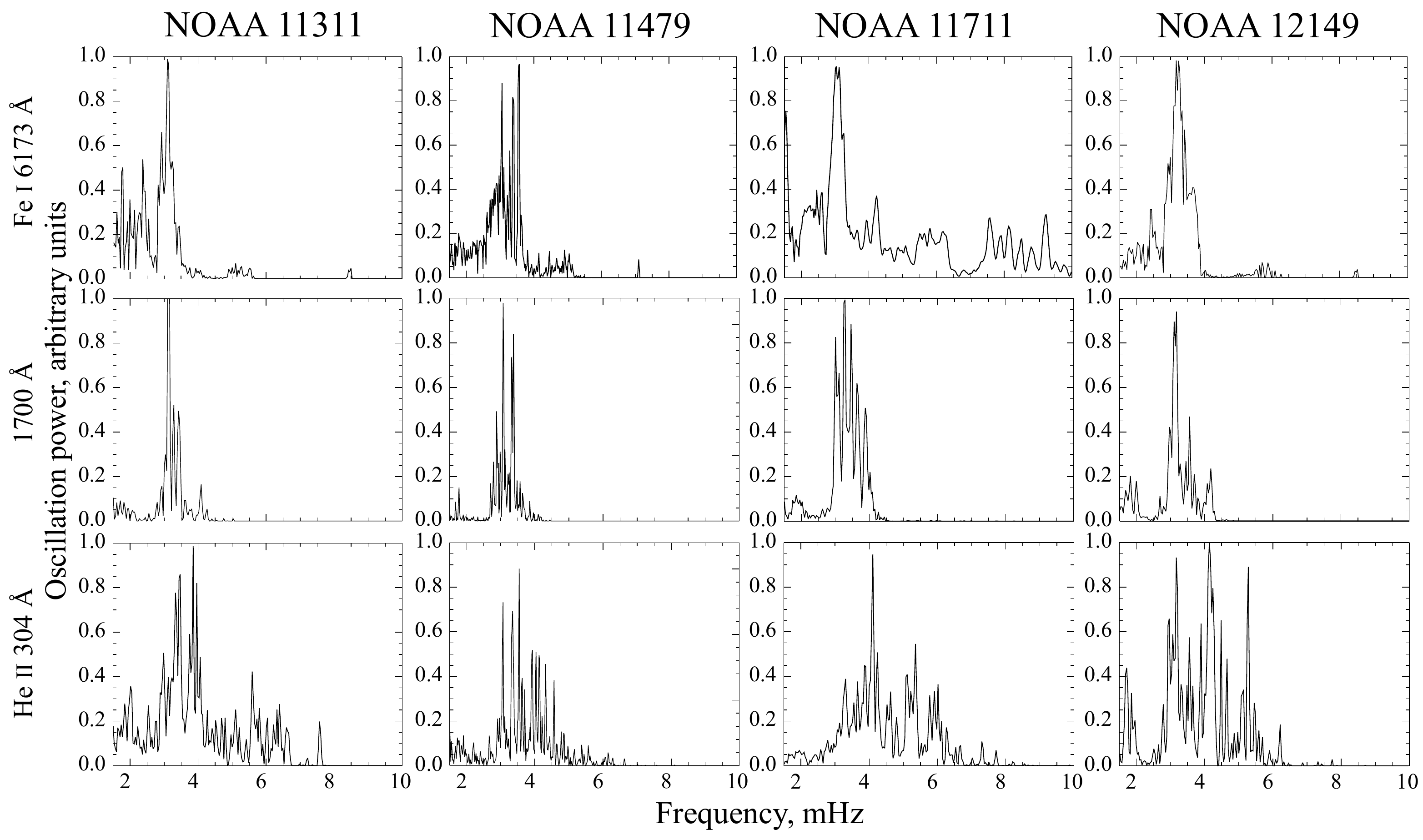}
}
\caption{Spectra azimuthally averaged over the region within the penumbrae marked
grey in Figure\,\ref{fig:radial}}
\label{fig:circularspec}
\end{figure}

 These plots behave as we expected---high-frequency range in the centre and
gradual decrease towards the sunspot boundaries. The interesting feature in all the
studied sunspots that caught our attention is the convergence of all the
profiles in the inner penumbrae (see Figure\,\ref{fig:radial}).
In case of NOAA\,11711 the profiles intersect in this region.
Figure\,\ref{fig:circularspec} shows intensity
oscillation power spectra azimuthally averaged over the region marked in
Figure\,\ref{fig:radial}. The panels for NOAA\,12149 show that in the transition
region (He\,\textsc{ii}\,304\,\AA) the highest peaks are shifted
to the higher frequencies. This can be explained by the inhomogeneity
of the penumbra or by the fact that sunspot has not purely circular shape.
There are five-minute oscillations
in the narrow region of the penumbra that dominate at all the heights
(Figure\,\ref{fig:circularspec}). We consider this region to be a
channel transporting five-minute oscillations from the photosphere up
through the chromosphere.

\par The interesting behaviour in this penumbra region
motivated a more detailed study of the distributions of several parameters
there, including those found in earlier works by other researchers.

\par
\begin{figure}[t]
\centerline{\includegraphics[width=13cm]{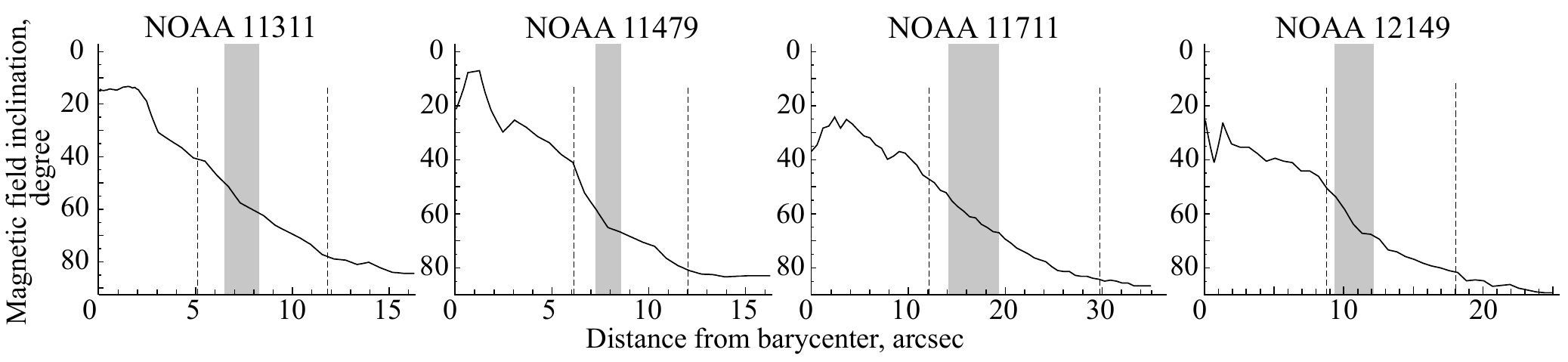}
}
\caption{Estimation of the magnetic field inclination at the He\,\textsc{ii}\,304\,\AA\ line
height based on the dominant frequency distributions. The region of interest is marked with grey area}
\label{fig:304mf}
\end{figure}

 Figure\,\ref{fig:304mf} shows the azimuthally averaged magnetic field
inclination at the He\,\textsc{ii}\,304\,\AA\ line formation level estimated
from the dominant frequency distributions. The details of the estimation
procedure can be found in \citet{Kobanov15}. The inner penumbra in these
distributions is characterized by the steepest inclination angle gradient; the
inclination angle there being 60--65\degree.

\par
\begin{figure}
\centerline{\mbox{\hskip1cm}
\includegraphics[width=8cm]{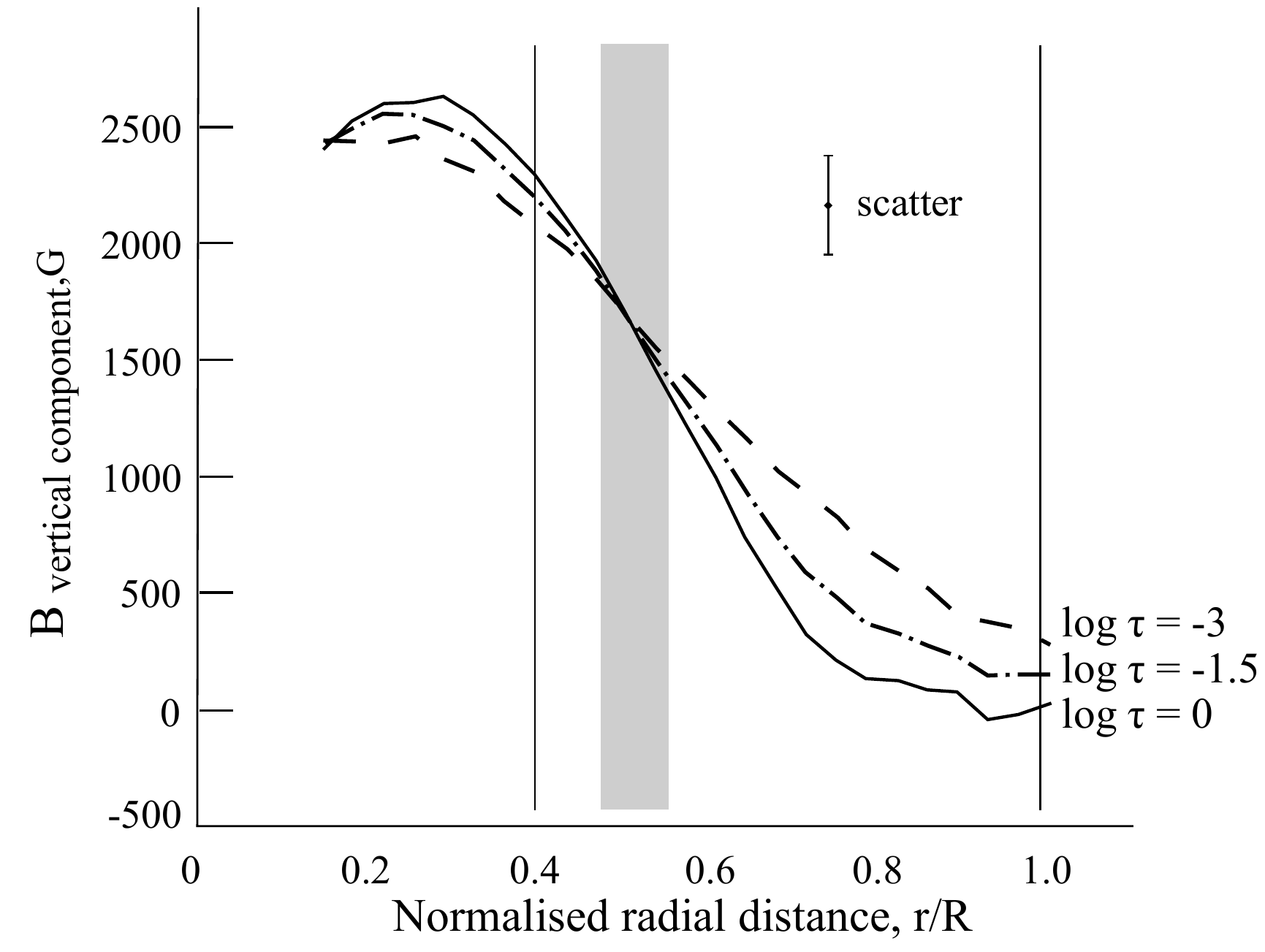}
}
\caption{Magnetic field vertical component averaged over
azimuth in sunspot NOAA 10923 in the deep photosphere at the continuum level, in
the mid-photosphere, and in the upper photosphere. The vertical lines mark the
penumbra boundaries. The region of interest is marked with grey area (courtesy
of Juan M. Borrero, see \citep{Borrero})}
\label{fig:BorreroMF}
\end{figure}

 Various signatures indicating  peculiar properties of the
inner penumbrae of sunspots have been found previously by other authors.
Identical field strength was found at three photospheric heights (deep
photosphere, log $\tau_{500} = 0$; mid photosphere, log $\tau_{500} = -1.5$;
and top of the photosphere, log $\tau_{500} = -3$) by \citet{Borrero}.
At about the same distance from the sunspot centre, the magnetic field profiles
of the three photospheric levels show the same value, and in the outer penumbra
the order of the profiles is reversed (Figure\,\ref{fig:BorreroMF}).
\cite{BellotRubio06}, see Figure~3 therein, noted a hump in the
inner penumbra in the azimuthally averaged temperatures at all the
photospheric heights ($-3 \leq \tau_{500} \leq 0$)
% (Figure\,\ref{fig:Temps}).
The amplitude of the hump was found to decrease with height. The
authors suggested that these temperature enhancements could be due to hot
penumbral tubes, by which plasma emerges from the sub-photospheric layers. Based
on these two works, one can conclude that the unique properties of the inner
penumbrae seem to originate in deeper levels than that we study here.

\par These tubes are probably related to the photospheric and chromospheric
Evershed flow, which peaks at the outer penumbra boundary and sharply terminates
at the same region of the inner penumbra %(see Figures\,\ref{fig:Vs},\,\ref{fig:Vs2},
(see Figure~8 in \cite{BellotRubio06}, and Figure~4 in \cite{Georgakilas03}).
Also, \citet{BellotRubio06} showed that
the microturbulence velocity rapidly drops to zero in the inner penumbra.

\par Today, we lack a comprehensive explanation for this phenomenon. Probably,
the key for understanding such a behaviour of the distributions is a model that
describes penumbra magnetic field as a series of two types of interlocking-comb
filaments \citep{Weiss}. The first one having a large inclination angle
are located at low heights and dive beneath the photosphere at the outer
penumbra boundary. These filaments are associated with the Evershed flow. The
second filament type is closer to vertical in orientation. Their magnetic field lines rise high in the
atmosphere and either return to the surface far from the spot or form an open
field line.

\par As follows from the aforesaid, interesting peculiarities are observed in
the behaviour of a number of physical parameters in the sunspots' inner
penumbra. In this paper, we hope to draw attention to this problem, solving
which requires widening the height range of data analysis and modelling.

%%%%%%%%%%%%%%%%%%%%%%%%%%%%%%%%%%%%%%%%%%%%%%%%%%%%%%%%%%%%%%%%%%%%%%%%%%%
{
\footnotesize \textbf{Acknowledgements}.
The study was performed with partial support of the Project No.\,16.3.2 of ISTP
SB RAS, by the Russian Foundation for
Basic Research under grants No. 15-32-20504 mol\_a\_ved and 16-32-00268 mol\_a.
We acknowledge the NASA/SDO science team for providing the data.
We are grateful to an anonymous referee for the helpful remarks and suggestions.
}

% \clearpage\newpage

%\acknowledgment US spelling: \verb+\acknowledgment+
%\acknowledgement British  spelling: \verb+\acknowledgement+

%%%%%%%%%%%%%%%%%%%%%%%%%%%%%%%%%%%%%%%%%%%%%%%%%%%%%%%%%%%%%%%%%%%%%%%%%%%
% \clearpage
\bibliographystyle{spr-mp-sola}
%\bibliographystyle{spr-mp-sola-cnd} %% Alternative style: no title,
                                      % no concluding page.

     % name your Bibtex file containing your references (.bib)
\tracingmacros=2
\clearpage
\bibliography{kolobov16}

% \end{article}

\end{document}